\documentstyle[pre,preprint,eqsecnum,aps]{revtex} 
\input epsf
\begin{document} 
\title{Critical End Point Behaviour in a Binary Fluid Mixture}

\author{Nigel B. Wilding \cite{ADDRESS}} 

\address{Institut f\"{u}r Physik, Johannes Gutenberg Universit\"{a}t,
\\ Staudinger Weg 7, D-55099 Mainz, Germany.}


\maketitle 
\begin{abstract}

We consider the liquid-gas phase boundary in a binary fluid mixture near its
critical end point.  Using general scaling arguments we show that the
diameter of the liquid-gas coexistence curve exhibits singular behaviour as
the critical end point is approached.  This prediction is tested by means of
extensive Monte-Carlo simulations of a symmetrical Lennard-Jones binary
mixture within the grand canonical ensemble.  The simulation results show
clear evidence for the proposed singularity, as well as confirming a
previously predicted singularity in the coexistence chemical potential
[Fisher and Upton, Phys.  Rev.  Lett.  {\bf 65}, 2402 (1990)].  The results
suggest that the observed singularities, particularly that in the
coexistence diameter, should also be detectable experimentally. 

\end{abstract}
\nopagebreak

\section{Introduction}
\label{sec:intro}

A critical end point occurs when a line of second order phase transitions
intersects and is truncated by a first order phase boundary, beyond which a
new {\em non-critical} phase is formed.  Critical end points are common
features in the phase diagrams of a many multicomponent systems such as
binary fluid mixtures, binary alloys and liquid crystals.  They also occur
under certain circumstances in pure systems having additional internal
degrees of freedom such as superfluids, as well as certain ferromagnets and
ferroelectrics. 

\begin{figure}[h]
\setlength{\epsfxsize}{14.0cm}
\centerline{\mbox{\epsffile{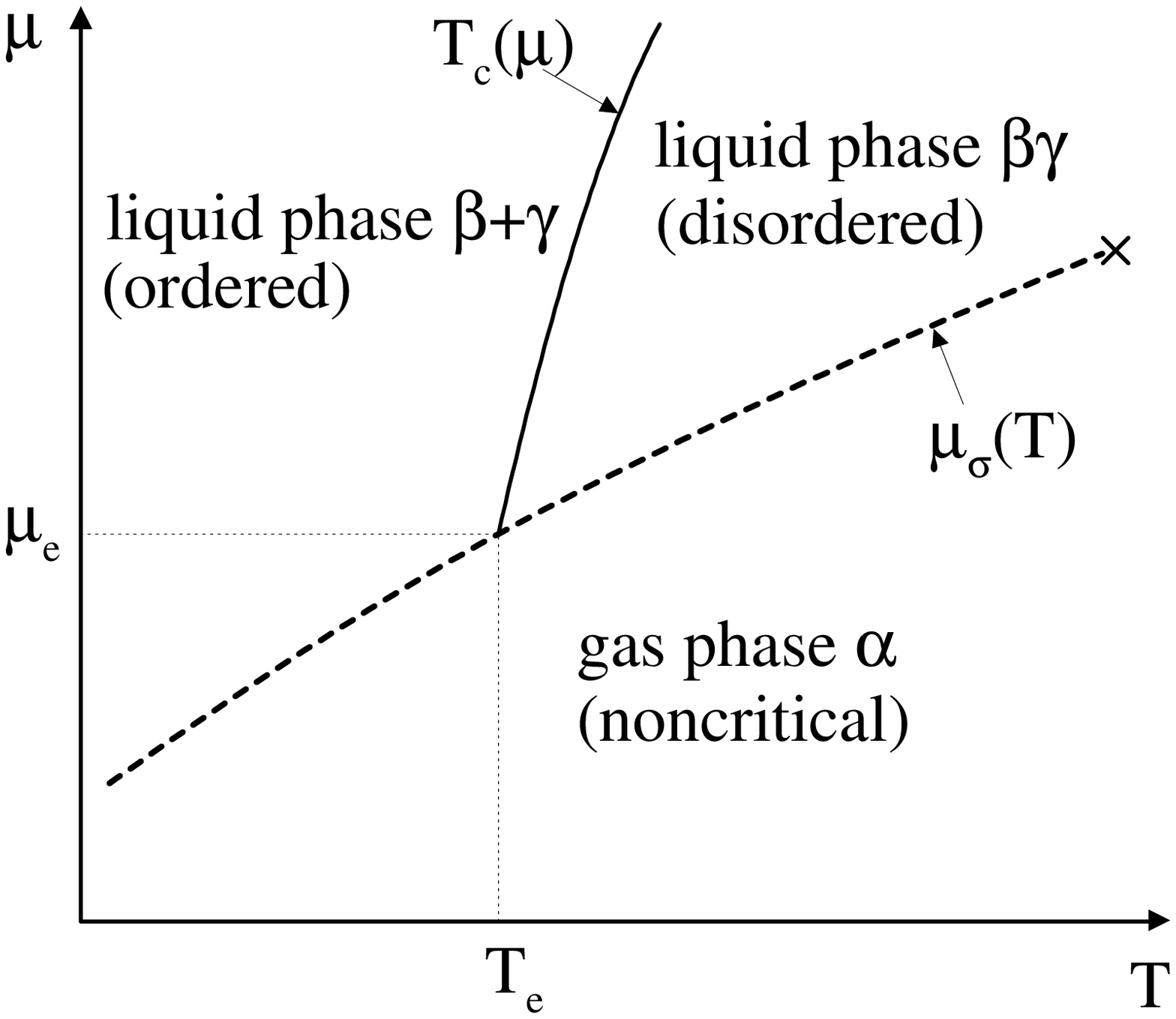}}} 

\caption{Schematic phase diagram of a binary fluid in the coexistence
surface $h=0$.  The broken line $\mu_\sigma(T)$ is the first order
liquid-gas phase boundary between the fluid and gas phase $\alpha$.  The
full line is the critical line of second order transitions $T_\lambda(\mu)$
separating the demixed phases $\beta$+$\gamma$, from the mixed phase
$\beta\gamma$.  The two lines intersect at the critical end point.}

\label{fig:schem}
\end{figure}

Perhaps the simplest system to exhibit a critical end point is a binary
fluid mixture.  Here the phase diagram is spanned by three
thermodynamic fields: the temperature $T$, a chemical potential $\mu$,
and an ordering field $h$ coupling to the concentration difference of
the two particle species. A number of different phase diagram
topologies are possible for binary fluids depending on the microscopic
interaction parameters, as systematically classified by Konynenburg and
Scott \cite{SCOTT}. The particular case on which we shall focus in this
work is of type II in their classification scheme, and is depicted
schematically in figure~\ref{fig:schem} for the subspace $h=0$.  Within
this subspace, two fluid phases $\beta$ and $\gamma$, each rich in one
of the two particle species, coexist with one another. By tuning $T$
and $\mu$, however, one finds a critical `$\lambda$' line, $T_c(\mu)$,
where both phases merge into a single disordered $\beta\gamma$ phase. 
The point at which the $\lambda$-line $T_c(\mu)$ intersects the first
order line of liquid-gas transitions $\mu_\sigma(T)$ marks the critical
end point ($T_e,\mu_e$). For $T<T_e$, the phase boundary
$\mu_\sigma(T)$ constitutes a triple line along which the fluid phases
$\beta$ and $\gamma$ coexist with the gas phase $\alpha$, while for
$T>T_e$, $\mu_\sigma(T)$ defines the region where the $\beta\gamma$ and
$\alpha$ phases coexist.  Precisely at the critical end point the
critical mixture of $\beta$ and $\gamma$ phases coexists with the gas
phase.  Since the gas phase does not participate in the criticality, it
is commonly referred to as a ``spectator'' phase. 

Interest in critical end points has recently been aroused following
theoretical work by Fisher and coworkers \cite{FISHER1}. On the basis of
phenomenological scaling and thermodynamic arguments, these authors argued
that the nonanalytic behaviour at the critical end point engenders a
free energy like singularity in the first order (spectator) phase boundary
$\mu_\sigma(T)$. The nature of this singularity, specifically its amplitude
ratio, was related to the universal features of the critical line.
Additionally, new universal amplitude ratios were proposed for the noncritical
surface tensions near the critical end point \cite{FISHER2}.  These
predictions were subsequently corroborated by analytical calculations on
extended spherical models \cite{BARBOSA,BARBOSA1}, as well as a Landau theory
study of a model ferroelectric \cite{HELENA}. To date, however, experimental
results pertaining to critical end points are scarce and those that do exist
have focused on the interfacial properties and the surface amplitude ratios
\cite{LAW,LAW1}. To the best of our knowledge there have been no reported
attempts to study the bulk coexistence properties and the matter of the
predicted singularity in the spectator phase boundary. There is a similar dearth
of simulation work on the subject, and we know of no detailed numerical
studies of critical end point behaviour, either in lattice or continuum
models.

In this paper we address the issue of the nature of the spectator phase
boundary near a critical end point by means of computer simulation of a
continuum binary fluid model \cite{NOTE0}.  Our paper is organised as
follows.  In section~\ref{sec:theory} we briefly review the principal
features of the scaling arguments of Fisher and coworkers, and show that in
addition to the previously predicted singularity in $\mu_\sigma(T)$, they
also imply singular behaviour in the {\em diameter} of the liquid-gas
coexistence curve at the critical end point.  In section ~\ref{sec:mc} we
detail extensive Monte-Carlo simulations of a symmetrical (Lennard-Jones)
binary fluid mixture.  We map the liquid-vapour coexistence curve and the
$\lambda$-line of the model by applying finite-size scaling analyses to
the probability distribution functions of appropriate observables.  The
results provide remarkably clear signatures of divergences in the
appropriate temperature derivatives of the coexistence diameter and the
phase boundary chemical potential, thus corroborating the theoretical
predictions.  Finally in section~\ref{sec:disc} we detail our conclusions. 

\section{The liquid-gas coexistence curve}
\label{sec:theory}

\subsection{The coexistence chemical potential}

In this subsection, we briefly review the scaling arguments of Fisher and
Barbosa \cite{FISHER1} concerning the singularity in the spectator phase
boundary at the critical end point. In so doing, we shall continue to employ
the language of the binary fluid mixture, although the arguments themselves
are not restricted to this case.

Within the grand canonical framework, liquid-gas coexistence is prescribed by
the equality of the Gibbs free energy $G=-k_BT\ln {\cal Z}$ in the respective
phases i.e. 
\begin{equation}
G_g(\mu_\sigma(T),T,h)=G_l(\mu_\sigma(T),T,h)
\label{eq:equal}
\end{equation}
Since the gas spectator phase is necessarily noncritical, its free energy
is analytic at the end point and can thus be expanded as

\begin{equation}
G_g(\mu,T,h)=G_e+G_1^g\Delta\mu+G_2^g t+G_3^g h +G_4^g\Delta \mu^2+\cdots
\label{eq:Ggas}
\end{equation}
where $t\equiv T-T_e$, and $\Delta\mu\equiv\mu-\mu_e$, with $\Delta\mu < 0$. 

The liquid phase on the other hand is critical and therefore contains both an
analytic (background) and a {\em singular} contribution to the free energy

\begin{equation}
G_l(\mu,T,h)=G_0(\mu,T,h)-|\tau|^{2-\alpha}{\cal G}_{\pm}(\hat{h}|\tau|^{-\Delta})
\label{eq:Gliq}
\end{equation}
where $G_0$ is the analytic part, while ${\cal G}^\pm(y)$ is a universal
scaling function which is a function of the relevant scaling fields
$\tau(T,\mu,h)=T-T_c(\mu)$ and $\hat{h}(T,\mu,h)\approx h$ that measure
deviations from the $\lambda$-line. ${\cal G}^\pm(y)$ must satisfy matching
conditions as $y\to\pm\infty$, and the quantities $\alpha$ and $\Delta$ are
respectively the specific heat and gap exponents associated with the
$\lambda$-line.

To linear order the scaling fields may be expanded as

\begin{mathletters}
\begin{eqnarray} 
\tau(T,\mu,h)       & = & t+ a_1 h+a_2 \Delta \mu \label{eq:tau}\\
\hat {h}(T,\mu,h)   & = & h+ b_1 t+b_2 \Delta \mu 
\end{eqnarray}
\end{mathletters}
where the $a_i$ and $b_i$ are non universal `field mixing' parameters
\cite{ANISIMOV,REHR} some of which have geometrical significance within the phase
diagram.  In particular, since $\tau=0$ specifies $T_c(\mu)$, one has at
the critical end point

\begin{equation}
a_2=-\frac{dT_c}{d\mu},
\end{equation} 
representing the gradient of the $\lambda$-line at the critical end point.

The critical free energy, equation~\ref{eq:Gliq}, can also be expanded in $\Delta
\mu, t$ and $h$. Recalling equation~\ref{eq:Ggas}, invoking
equation~\ref{eq:equal} and solving for $\mu(T)$ then yields \cite{FISHER1}

\begin{equation}
\mu_\sigma(T)-\mu_o(T)\approx -X_\pm|t|^{2-\alpha}-Y_\pm|t|^\beta
|h|-\frac{1}{2}Z_\pm|t|^{-\gamma} h^2, 
\label{eq:csing}
\end{equation}
valid as $ T\to T_e\pm ,\, h\to 0$. Here $X_\pm, Y_\pm, Z_\pm$ are critical
amplitudes, and $\alpha, \beta, \gamma$ are the usual critical exponents
characterising the $\lambda$-line.  Corrections to scaling have been neglected
and the background term has the expansion

\begin{equation}
\mu_o(T)=\mu_e+g_1t+g_2h+\cdots
\label{eq:moexp}
\end{equation}

If we restrict out attention to the coexistence surface $h=0$ and assume
$\alpha>0$, equation~\ref{eq:csing} implies a specific heat-like divergence in
the {\em curvature} of the spectator phase boundary

\begin{equation}
\frac{d^2\mu_\sigma}{dT^2}\approx -\hat{X}_\pm|t|^{-\alpha}  
\label{eq:cdiv}
\end{equation}
where the amplitude ratio $\hat{X}_+/\hat{X}_-$ is expected to be
universal \cite{FISHER1}. 

\subsection{The coexistence diameter}

Let us now examine the behaviour of the coexistence particle number density
in the vicinity of the critical end point.  Specifically, we shall focus on
the coexistence diameter in the region $\hat{h}=0$, defined by

\begin{equation}
\rho_d(T)\equiv\frac{1}{2}[\rho_g(\mu_\sigma(T))+\rho_l(\mu_\sigma(T))].
\end{equation}
The particle density is obtainable from the Gibbs potential as

\begin{equation}
\rho=-\frac{1}{V}\left ( \frac{\partial G}{\partial\mu}\right )_{T,h}
\label{eq:rho}
\end{equation}
so that

\begin{equation}
\rho_d(T)=
-\frac{1}{V}\left (\frac{\partial G_\sigma(\mu_\sigma(T),T)}{\partial\mu}\right )_T
\label{eq:rho_d}
\end{equation}
with 

\begin{equation}
G_\sigma(\mu_\sigma(T),T)=[G_g(\mu_\sigma(T),T)+G_l(\mu_\sigma(T),T)]/2.
\end{equation}

Appealing to eqs.~\ref{eq:Ggas} and \ref{eq:Gliq}, and noting that
$\beta=2-\alpha-\Delta$, then yields for the singular behaviour

\begin{eqnarray}
\rho_d(T)  & = &U_\pm|\tau(T,\mu_\sigma(T))|^\beta + V_\pm|\tau(T,\mu_\sigma(T))|^{1-\alpha}\nonumber\\
 & & + \hspace{0.7mm} \mbox{terms analytic at} \hspace{1mm} T_e
\label{eq:rhod}
\end{eqnarray}
as $\tau\to 0$. Here the non-universal critical amplitudes take the form

\begin{mathletters}
\begin{eqnarray}
\label{eq:amp1}
U_\pm &=& b_2 {\cal G}_\pm^\prime (0)\\
\label{eq:amp2}
V_\pm &=& a_2 (2-\alpha){\cal G}_\pm(0)
\end{eqnarray}
\end{mathletters}
with ${\cal G}_\pm^\prime(z) =d{\cal G}_\pm/dz$.

Now, along the liquid-gas coexistence curve, one has from equation~\ref{eq:tau}

\begin{equation}
|\tau(T,\mu_\sigma(T))|=|t+a_2(\mu_\sigma(T)-\mu_e)|.
\end{equation} 
Recalling equations~\ref{eq:csing} and~\ref{eq:moexp} (and setting $h=0$), 
one then finds

\begin{equation}
|\tau(T,\mu_\sigma(T))|=|t|[1+a_2g_1+O(|t|^{1-\alpha})].
\end{equation}
Thus to leading order one can write

\begin{eqnarray}
\rho_d(T)  & = & \tilde U_\pm|t|^\beta + \tilde V_\pm|t|^{1-\alpha}\nonumber\\
 & & + \hspace{0.7mm} \mbox{terms analytic at} \hspace{1mm} T_e,
\label{eq:rhod1}
\end{eqnarray}
where $\tilde U_\pm=(1+a_2g_1)^\beta U_\pm$ and $\tilde
V_\pm=(1+a_2g_1)^{1-\alpha} V_\pm$.  We note that this expression is of the
same form as the singularity in the overall density on the $\lambda-$line
\cite{ANISIMOV}.

A special case of equation~\ref{eq:rhod1}, relevant to the present work, is
that for a symmetrical fluid having energetic invariance under $h\to-h$.  In
this case one finds on symmetry grounds that the field mixing parameters
$b_1=b_2=0$ and hence from equation~\ref{eq:amp1}

\begin{equation}
\rho_d(T) = \tilde V_\pm|t|^{1-\alpha} + \mbox{terms analytic at} \hspace{1mm} T_e,
\label{eq:rhod2}
\end{equation}
which implies a divergent diameter derivative

\begin{equation}
\frac{d\rho_d(T)}{dT} \approx \hat{V}_\pm|t|^{-\alpha}
\label{eq:ddiv}
\end{equation}
where $\hat{V}_\pm=(1-\alpha)\tilde V_\pm$. Since this divergence occurs in the first derivative of the observable
$\rho_d(T)$, it is in principle more readily visible than that in the second
derivative of $\mu_\sigma(T)$, cf.  eq.~\ref{eq:cdiv}.  As we shall now show,
however, clear signatures of both divergences are readily demonstrable by
Monte-Carlo simulation.

\section{Monte Carlo simulations}

\label{sec:mc}

\subsection{Model and algorithmic considerations} 
\label{sec:model}

The simulations described here were performed for a symmetrical binary fluid
model using a Metropolis algorithm within the grand canonical ensemble (GCE)
\cite{ALLEN}. The fluid is assumed to be contained in volume $V=L^3$ with
periodic boundary conditions. The grand-canonical partition function takes the
form

\begin{equation}
\label{eq:zdef}
{\cal Z}_{L}  = \sum _{N_1=0}^{\infty }\sum _{N_2=0}^{\infty }\prod _{i=1}^{N} \left\{\int d\vec{r}_i\right\}
e^{ \left[ \mu N
-\Phi (\{ \vec{r} \} ) + h(N_1-N_2) \right]}
\end{equation}
where $\Phi=\sum_{i<j}\phi(r_{ij})$ is the total configurational energy, $\mu$
is the chemical potential, and $h$ is the ordering field (all in units of
$k_BT$).  $N=N_1+N_2$ is the total number of particles of types $1$ and
$2$. 

The interaction potential between particles $i$ and $j$ was assigned the
familiar Lennard--Jones (LJ) form

\begin{equation} 
\phi(r_{ij})=4\epsilon_{mn}[(\sigma/r_{ij})^{12}-(\sigma/r_{ij})^6],
\label{eq:LJdef} 
\end{equation} 
where $\sigma$ is a parameter which serves to set the interaction range, while
$\epsilon_{mn}$ measures the well-depth for interactions between particles of
types $m$ and $n$. In common with most other simulations of Lennard-Jones
systems, the potential was truncated at a cutoff radius $r_c=2.5\sigma$
(irrespective of the species of the interacting particles) and left unshifted.
To simplify the task of locating interacting particles, a cell decomposition
scheme was employed in which the total system volume was partitioned into
cubic cells of side $r_c$. Interaction emanating from a specific particle then
only extend as far as the $26$ neighbouring cells.

An Ising like symmetry was imposed on the model by choosing
$\epsilon_{11}=\epsilon_{22}=\epsilon>0$.  This choice endows the system with
energetic invariance under $h\to -h$, thereby ensuring that the critical end
point lies in the surface $h=0$. We shall accordingly restrict our attention
henceforth to this regime.  A further parameter $\epsilon_{12}=\delta$ was
used to control interactions between unlike particles.  The phase diagram of
the model in the surface $h=0$ is then uniquely parameterised by the ratio
$\delta/\epsilon$. Physically, the role of this ratio in determining the
form of the phase diagram can be understood as follows. For
$\delta/\epsilon\lesssim 1$, the energy penalty associated with contacts
between unlike particles is small and hence there is little incentive for
phase separation unless the temperature is very low and the density very
high. One therefore expects a phase diagram having a critical end point
temperature $T_e\ll T^{lg}_c$ and density $\rho_e\gg\rho^{lg}_c$, where
$T^{lg}_c$ and $\rho^{lg}_c$ are the liquid-gas critical temperature and
density respectively.  Choosing a smaller value of $\delta/\epsilon$, however,
moves the end point towards the liquid-gas critical point, into which it
merges for sufficiently small $\delta/\epsilon$, forming a {\em tricritical
point} \cite{WILDING0}.  Empirically we find that for $\delta/\epsilon=0.6$,
there is a tricritical point, while for $\delta/\epsilon=0.75$ there is a
critical end point having $\rho_e\approx 2.3\rho^{lg}_c$.  The location of the
critical end point for larger values of $\delta/\epsilon$ could not be
reliably determined since its density is above that at which the grand
canonical particle insertion algorithm (see below) is operable.  In this work,
all simulations were performed with $\delta/\epsilon=0.7$, which yields
critical end point parameters $T_e\approx 0.93 T^{lg}_c$, $\rho_e\approx
1.75\rho^{lg}_c$.  This temperature is sufficiently small compared to
$T^{lg}_c$ that critical density fluctuations don't obscure the end point
behaviour, while at the same time $\rho_e$ is not so large as to unduely
hinder particle insertions.

In order to sample ergodically the phase space of the model, two sorts of
Monte-Carlo action are necessary. The first is a particle transfer step in
which one attempts either to insert a particle at a randomly chosen position,
or alternatively, to delete a randomly chosen existing particle.  Candidate
particles for insertion are assigned a species type ($1$ or $2$) with equal
probability.  The second sort of action is an identity swop, in which one
chooses an existing particle at random and attempts to change its identity
($1\to2$ or $2\to 1$). Combined use of these operations samples the grand
canonical ensemble in which the particle density $\rho=N/V$, energy density
$u=\Phi/V$ and number difference density $m=(N_1-N_2)/V$ all fluctuate.

In accordance with convention, we shall employ dimensionless units to
express our data:

\begin{equation}
\tilde{\rho} = \rho \sigma ^3, \hspace{8mm} \tilde m=m\sigma^3, \hspace{8mm}\tilde u=u\sigma^3 \\
\label{eq:rhostar}
\end{equation}

\begin{equation}
\tilde{T} = k_BT/\epsilon 
\end{equation}
We also note for future reference that our algorithm utilises not the
true chemical potential $\mu$, but an effective chemical potential
$\tilde{\mu}$ to which the true chemical potential is related by

\begin{equation}
\mu = \tilde{\mu} + \mu _0 - \ln \left ( N/L^3 \right ) \label{eq:mustar}
\end{equation}
where $\mu_0$ is the chemical potential in the non-interacting (ideal
gas) limit. It is this effective value that features in the results
that follows.

\subsection{Method and Results}
\label{sec:results}

In the course of the simulations, three systems sizes of volume
$V=(7.5\sigma)^3$, $V=(10\sigma)^3$ and $V=(12.5\sigma)^3$ were studied,
corresponding to average particle numbers $N\approx 250$, $N\approx 600$
and $N\approx 1200$ respectively at the critical end point (whose location we
discuss below).  Following equilibration, runs comprising up to $6\times 10^9$
MCS \cite{NOTE3} were performed and the joint distribution
$p_L(\tilde\rho,\tilde m,\tilde u)$ was sampled approximately every $10^4$ MCS and
accumulated in the form of a histogram. For each $L$, simulations were carried
out at several (typically $5$) temperatures along the liquid-gas coexistence curve. 

In order to locate the liquid-gas coexistence curve, the finite-size form of
the density distribution $p_L(\tilde\rho)$ was studied.  Precisely at
coexistence, $p_L(\tilde\rho)$ is (to within corrections exponentially small
in $L$) double peaked with equal weight in both peaks
\cite{BORGS}.  For a given simulation temperature, this `equal weight'
criterion can be used to determine the coexistence chemical potential to high
accuracy. Close to the liquid-gas critical point i.e. when $T\lesssim
T_c^{lg}$, this task is straightforward; one simple tunes the chemical
potential at constant $T$ until the measured form of $p_L(\tilde\rho)$ is
double peaked. By contrast, however, in the strongly first order regime ($T\ll
T_c^{lg}$), this approach cannot be used to obtain the coexistence form of
$p_L(\tilde\rho)$ due to the large free energy barrier separating the
coexisting phases. This barrier leads to metastability effects and
prohibitively long correlation times.  To circumvent this difficulty one must
employ advanced sampling schemes. One such scheme is the multicanonical
preweighting method \cite{BERG} which encourages the simulation to sample the
interfacial configurations of intrinsically low probability. This is achieved
by incorporating a suitably chosen weight function in the Monte-Carlo update
probabilities. The weights are subsequently `folded out' from the sampled
distribution to yield the correct Boltzmann distributed quantities. Use of
this method permits the direct measurement of the order parameter probability
distribution at first order phase transitions, even when the distribution
spans many decades of probability.

To fully capitalise on the data gathered in the simulations, the histogram
reweighting technique \cite{FERRENBERG} was employed. This technique rests on
the observation that histograms accumulated at one set of model parameters can
be reweighted to yield estimates of histograms appropriate to another set of
not-too-distant parameters. Results from individual simulation runs at different
parameters can also be combined in a systematic fashion to provide information
over larger regions of the phase diagram. When used in tandem with
multicanonical preweighting, the histogram reweighting technique constitutes a
powerful tool for mapping the coexistence curve properties of continuum fluid
models \cite{WILDING1}.

\begin{figure}[h]
\setlength{\epsfxsize}{11.0cm}
\centerline{\mbox{\epsffile{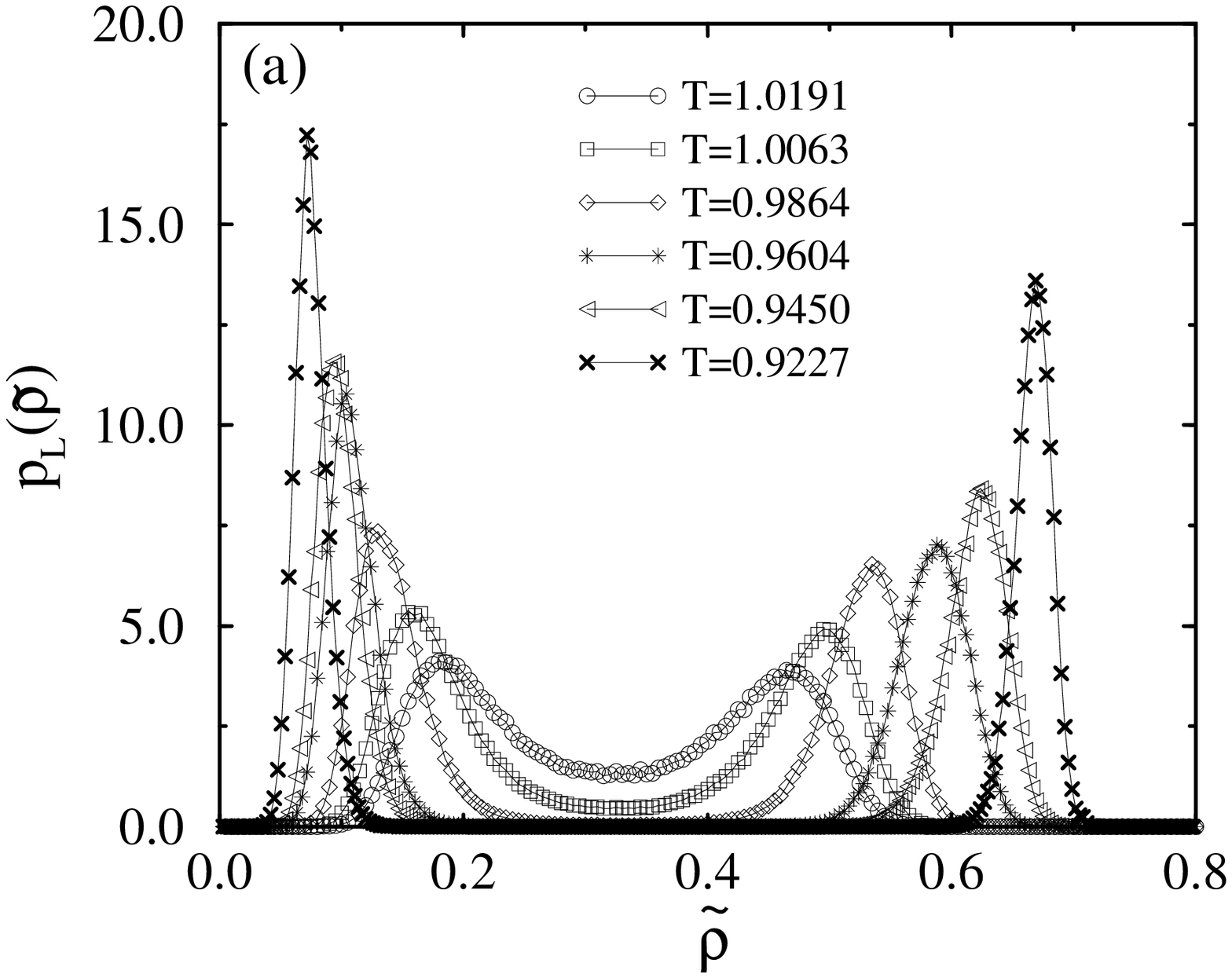}}} 
\centerline{\mbox{\epsffile{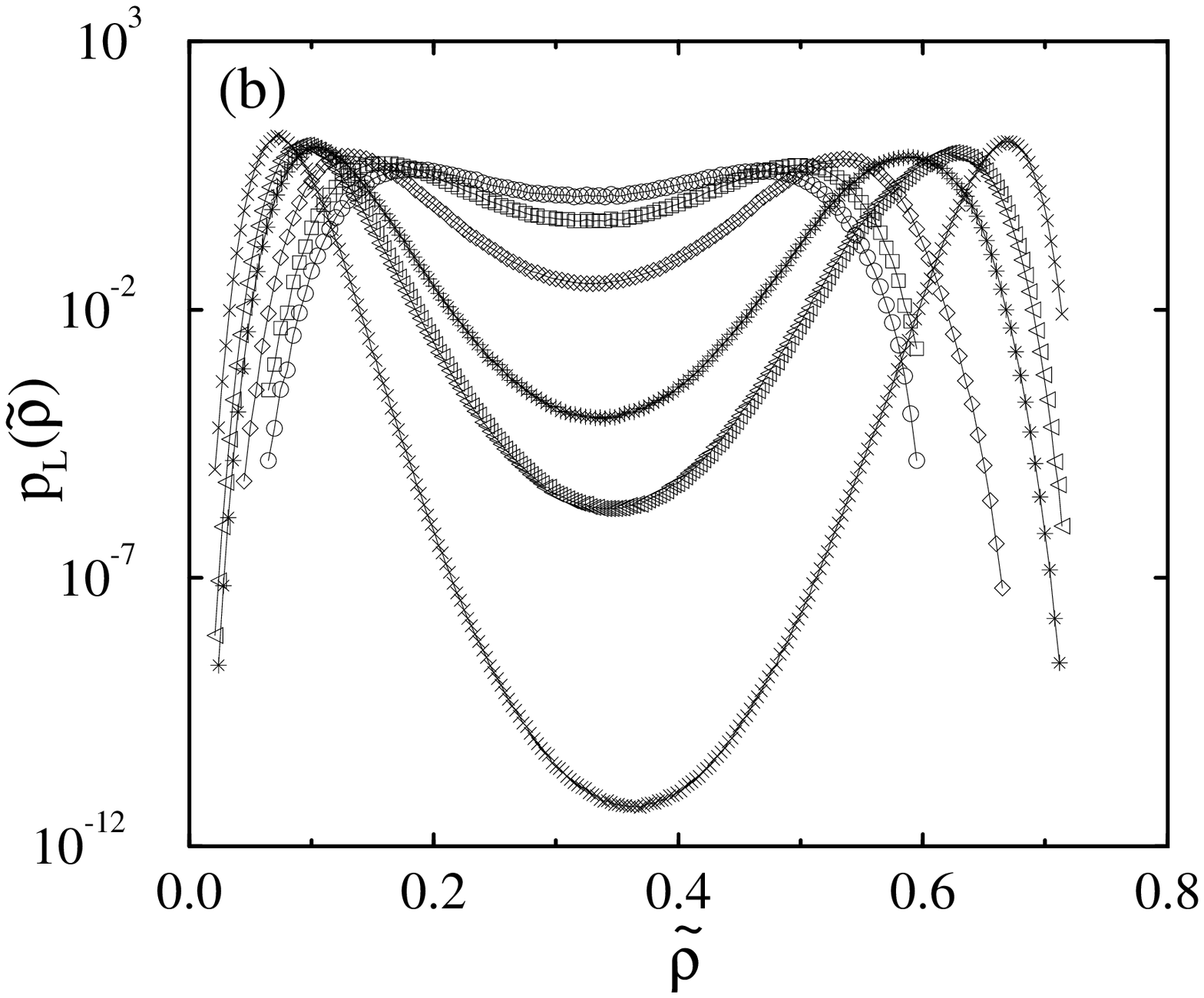}}} 

\caption{{\bf (a)}. Estimates of the coexistence density distributions
for the $L=10\sigma$ systems size, for a range of subcritical temperatures,
obtained as described in the text. The lines are merely guides to the
eye. Statistical errors do not exceed the symbol sizes. {\bf (b)}. The same data
expressed on a log scale.}

\label{fig:cxdists}
\end{figure}

\begin{figure}[h]
\setlength{\epsfxsize}{14.0cm}
\centerline{\mbox{\epsffile{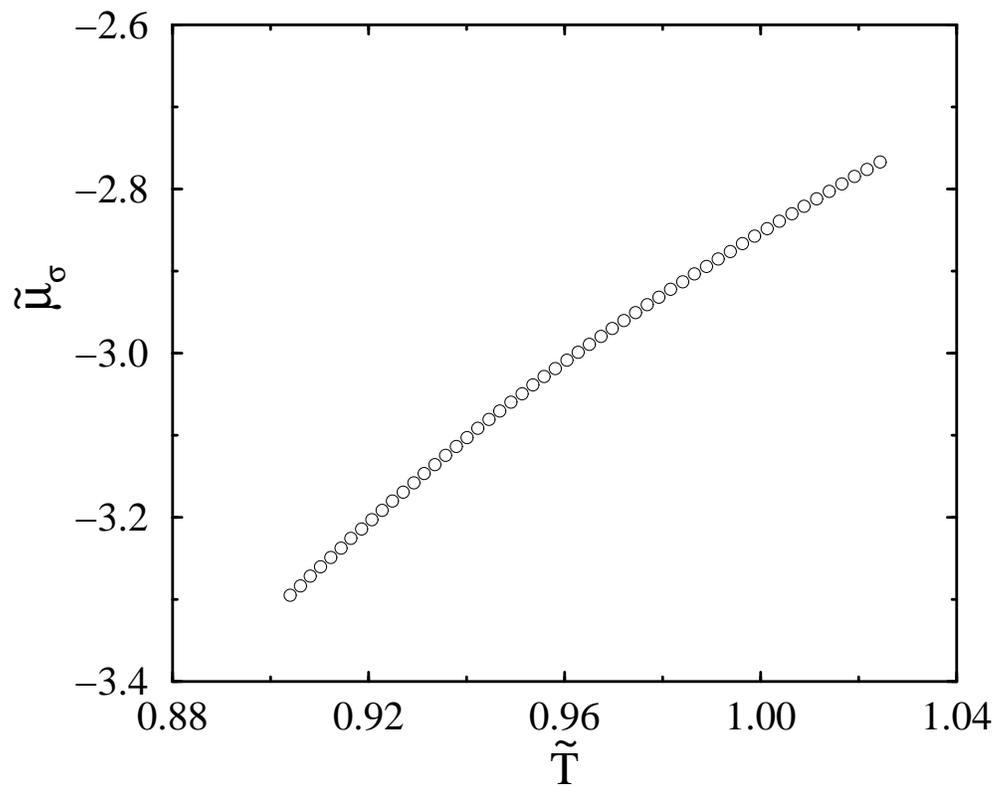}}} 

\caption{The temperature dependence of the coexistence chemical potential
$\tilde\mu$ obtained from the histogram reweighting of the distributions shown
in figure~\protect\ref{fig:cxdists}. Statistical errors do not exceed the
symbol sizes.}

\label{fig:muT}
\end{figure}

\begin{figure}[h]
\setlength{\epsfxsize}{14.0cm}
\centerline{\mbox{\epsffile{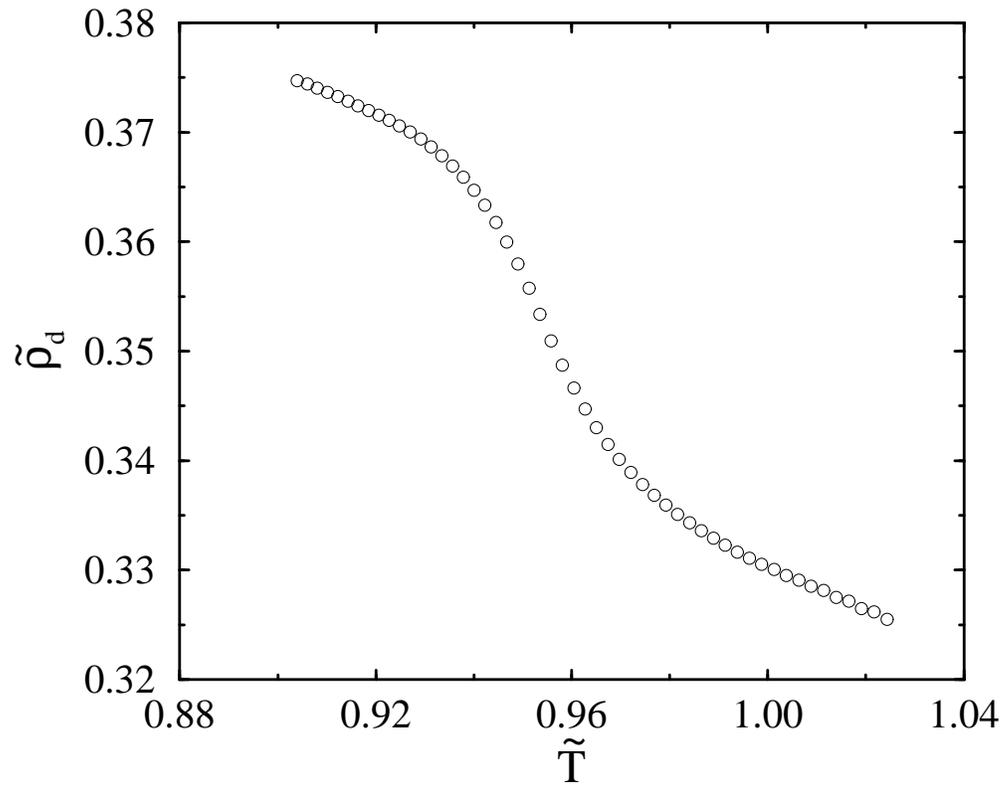}}} 

\caption{The temperature dependence of the coexistence curve diameter $\tilde\rho_d=(\tilde\rho_g+\tilde\rho_l)/2$,  as obtained from histogram reweighting of the
distributions shown in figure~\protect\ref{fig:cxdists}. Statistical errors do
not exceed the symbol sizes.}

\label{fig:rhod}
\end{figure}

Using the multicanonical preweighting technique, simulations of the
liquid-gas coexistence curve of the binary fluid were performed.  To begin
with, the region near the liquid-gas critical point was studied since there
the free energy barrier between the coexisting phases is small and no preweighting
function is need.  Thereafter, the temperature was reduced in a stepwise
fashion and the coexistence form of $p_L(\tilde\rho)$ was collected.  For
each successively lower temperature studied, histogram reweighting was used
to obtain a suitable preweighting function by extrapolating from the
coexistence histograms previously obtained at higher temperatures.  Further
details concerning this strategy have been given elsewhere \cite{WILDING1}. 
Figure~\ref{fig:cxdists}(a) shows the resulting coexistence forms of
$p_L(\tilde\rho)$ for the system of size $L=10\sigma$, at a selection of
temperatures in the range $0.9 \tilde T_C^{lg}<\tilde T<\tilde T_c^{lg}$. 
Also shown in figure~\ref{fig:cxdists}(b) is the same data expressed on a
log scale.  The corresponding estimates for the coexistence chemical
potential and the coexistence diameter are plotted in figures~\ref{fig:muT}
and ~\ref{fig:rhod} respectively.  The gas and liquid densities required for
the diameter calculation were obtained as the average densities of the
respective peaks of $p_L(\tilde\rho)$.  It is interesting to note from
figure~\ref{fig:cxdists}(b), that for the lowest temperature studied
($T\approx 0.9T_c$), the ratio of minimum to maximum in the distribution is
approximately $10^{12}$.  A free energy barrier of this magnitude would, of
course, represent an insurmountable obstacle to phase space evolution, were
it not for the use of multicanonical preweighting. 

\begin{figure}[h]
\setlength{\epsfxsize}{14.0cm}
\centerline{\mbox{\epsffile{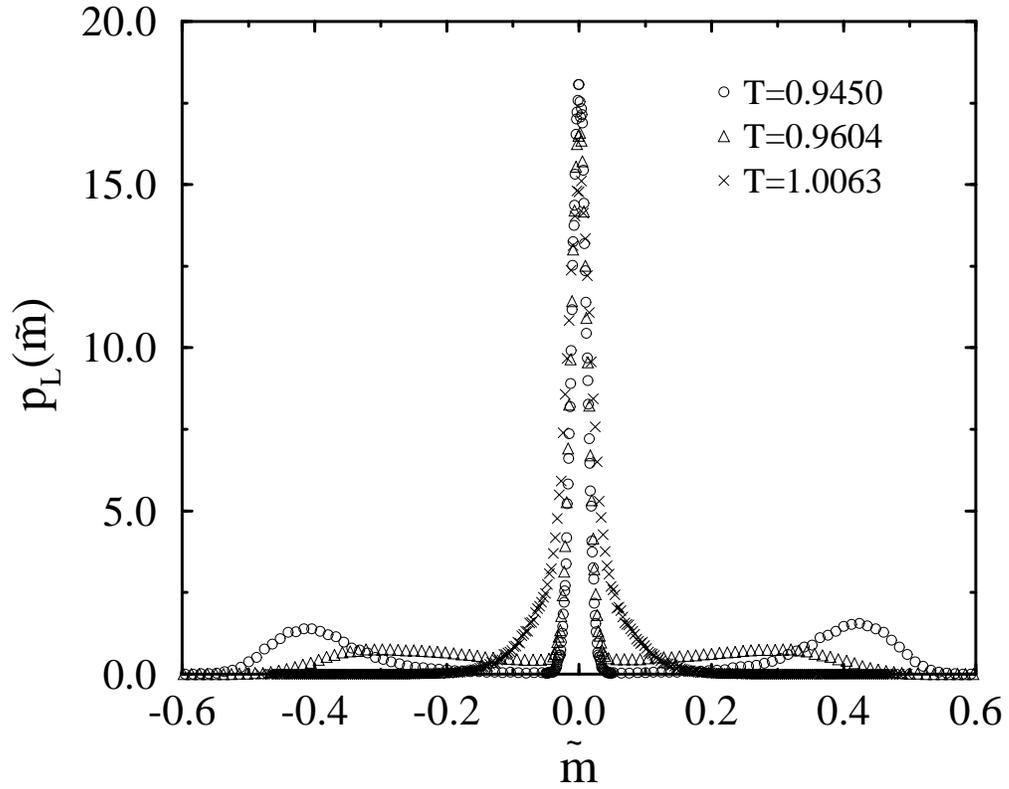}}} 

\caption{Estimates of the the number difference order parameter
distribution $p_L({\tilde m})$ for the $L=10\sigma$ system size at
three temperatures spanning $T_e$.}

\label{fig:magdists}
\end{figure}

It is also instructive to examine the coexistence behaviour of the 
number difference order parameter distribution $p_L(m)$, for
temperatures above and below $T_e$, as shown in 
figure~\ref{fig:magdists}. Well below $T_e$, this distribution is
three-peaked, with one narrow peak centered on $\tilde m=0$ and two
broader peaks centered on positive and negative values of $\tilde m$.
The peak at $\tilde m=0$ corresponds to the disordered gas phase, while
the degenerate peaks at positive and negative values of $m$ represent
the ordered $A$-rich and $B$-rich liquid phases respectively. As one
approaches the critical end point, however, the liquid peaks become
broader and overlap with the gas phase peak, ultimately leaving just
one single peak at $\tilde m=0$ for $T>T_e$.

To determine the locus of the $\lambda$-line and the critical end point
in which it formally terminates, finite-size scaling (FSS) methods were
employed. On the $\lambda$-line, criticality is signalled by the
asymptotic scale invariance of $p_L(\tilde m)$.  A useful dimensionless
measure of the form of this distribution is its fourth order cumulant
ratio $U_L=1-3\langle \tilde m^4\rangle/\langle \tilde m^2\rangle^2$
\cite{BINDER}.  If one plots $U_L(\tilde T)$  for a given constant
$\tilde\mu$, one expects the curves for each $L$ to intersect at the
critical de-mixing temperature.  For binary fluids with short ranged
interactions, the critical behaviour on the $\lambda$-line and the
critical end point \cite{ZIMAN} is expected to be Ising like.  One thus
expects that asymptotically (for sufficiently large $L$) there will be
a unique crossing point at a universal value $U_L^\star$. Ising model
simulations \cite{HILFER} give the estimate $U_L^\star= 0.470(3)$.

\begin{figure}[h]
\setlength{\epsfxsize}{12.0cm}
\centerline{\mbox{\epsffile{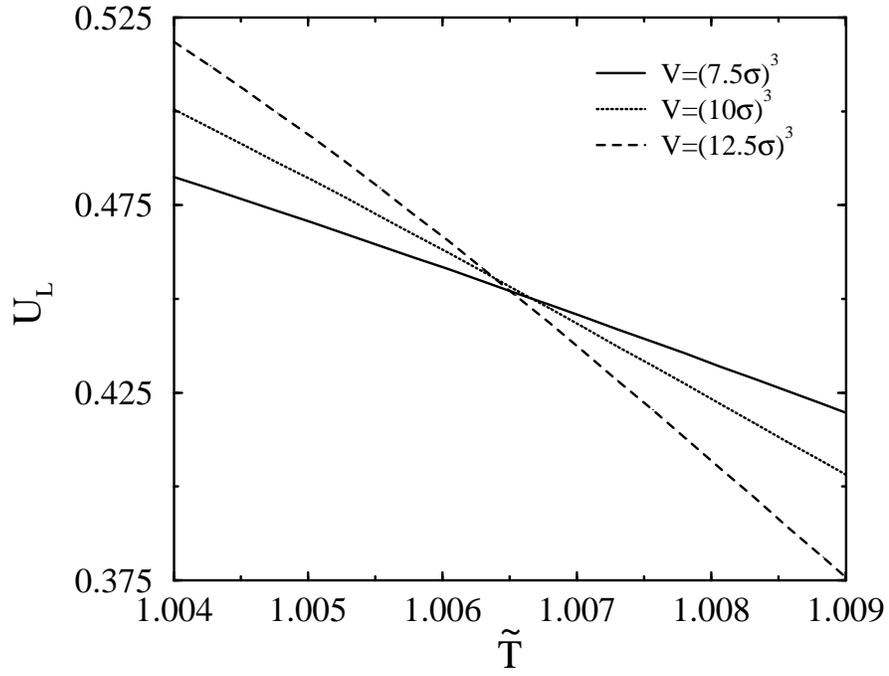}}} 

\caption{The measured cumulant ratio $U_L(\tilde T)$ for
the three system sizes studied. The lines are the results of histogram
reweighting of single simulations performed at $\tilde T=1.005,
\tilde\mu=-2.95$}

\label{fig:ullam}
\end{figure}

We have measured $U_L(\tilde T)$ for several value of the chemical potential $\tilde
\mu$, thus enabling an estimate of the locus of the $\lambda$-line. For each
chosen value of $\tilde\mu$, an initial rough estimate of the critical
temperature was obtained from a number of short simulations, in which the form
of $p_L(\tilde m)$ was observed visually. Longer runs at this temperature were then
carried out for each system size in order to facilitate a more precise
determination of $\tilde T_c$.  A representative plot of $U_L$ as a function
of $\tilde T$ and $L$ is shown in figure~\ref{fig:ullam} for the case
$\tilde\mu=-2.95$.  Each of the curves shown represent the histogram
extrapolation of data obtained from a single simulation at the temperature
$\tilde T=1.005$.  Evidently there is no unique cumulant intersection point
for the three system sizes, rather the crossing point for the $L=7.5\sigma$
and $L=10\sigma$ system sizes occurs at larger $\tilde{T}$ and smaller $U_L$
than for the $L=10\sigma$ and $L=12.5\sigma$ system sizes.  It is reasonable
to assume, however, that for sufficiently large $L$, the cumulant intersection
point {\em would} converge on the universal value, $U_L^\star= 0.470(3)$,
but that for the system sizes studied here this limit is not attained due to
corrections to finite-size scaling and (to a lesser extent) statistical
error. In view of this, we adopt as our estimate of $\tilde T_c(\tilde\mu)$,
that temperature at which the cumulant intersection occurs for the two largest
system sizes.

\begin{figure}[h]
\setlength{\epsfxsize}{12.0cm}
\centerline{\mbox{\epsffile{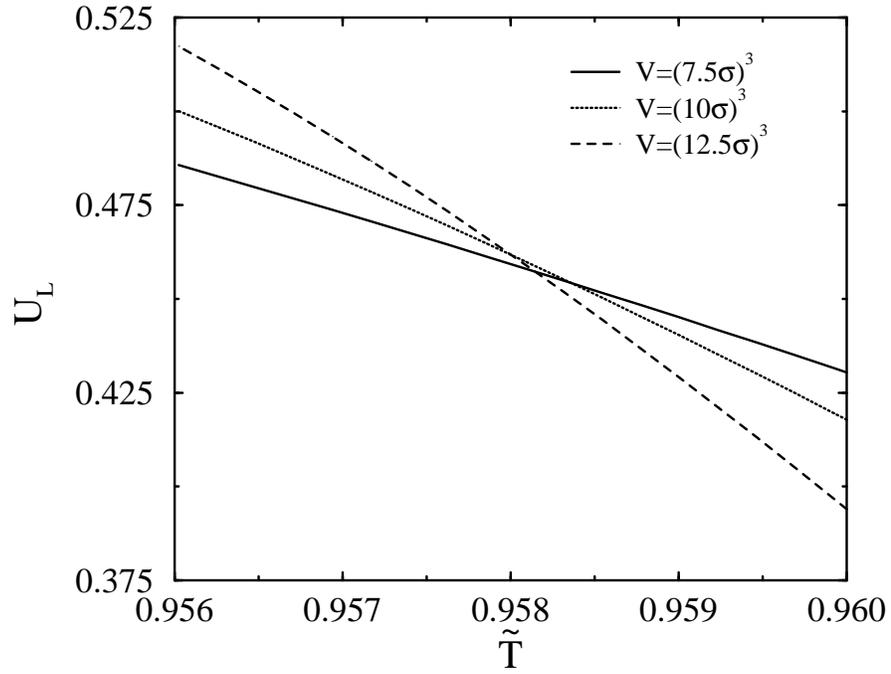}}} 

\caption{The cumulant ratio $U_L$ for the distribution $p_L(\tilde m)$,
obtained according to the procedure described in the text. The lines are the
results of histogram reweighting of liquid-phase simulations performed at
liquid-gas coexistence.}

\label{fig:ulend}
\end{figure}

The critical end point was estimated in a similar manner, by studying the
cumulant ratio $U_L$ for $p_L(\tilde m)$ as a function of $\tilde T$ and $L$
along the {\em liquid} branch of the coexistence curve.  In principle, the
form of $p_L(\tilde m)$ corresponding to the liquid phase can be extracted
from the joint coexistence distribution $p_L(\tilde\rho,\tilde m,\tilde u)$,
since this embodies contributions from both phases.  Assuming there is a
clear distinction between the gas and liquid phases (ie.  when $T\ll
T_c^{lg}$), one can isolate the liquid branch component simply by
disregarding all those histogram entries having $\tilde\rho<\tilde\rho_d$. 
However, it turns out to be difficult to obtain sufficient statistics for
$p_L(\tilde m)$ in this way.  This is because the preweighted simulation
uniformly samples {\em all} densities between the gas and liquid phases. 
When the gas and liquid densities are well separated, the simulation spends
only a relatively small fraction of the time sampling liquid-like
configurations and accordingly the statistical quality of the histogram for
$p_L(\tilde m)$ is low.  To circumvent this problem, it is expedient to
perform a simulation that samples {\em only} the liquid phase, while
neverthless remaining at coexistence.  This is achievable by setting $\tilde
T$ and $\tilde\mu$ to their coexistence values and performing a simulation
{\em without} preweighting the density distribution.  If a starting
configuration having a liquid-like density is chosen, the simulation will
then remain in the liquid phase by virtue of the large free energy barrier
separating it from the gas phase.  The results of implementing this
procedure are shown in figure~\ref{fig:ulend}, in which $U_L$ is plotted as
a function of temperature along the liquid branch of the coexistence curve
in the neighbourhood of the critical end point.  The curves exhibit an
intersection at $\tilde{T}_e=0.958(2)$, with a cumulant value $U_L=0.47$ for
the two largest system sizes and $U_L=0.455$ for the smallest two system
sizes.  The associated estimate of the critical end point chemical potential
and density are $\tilde\mu_e=-3.017(3), \tilde\rho_e=0.587(5)$. 

In the interests of completeness we have also obtained the location of the
liquid-gas critical point. This was achieved via a FSS study of the
density-like ordering operator. Details of the approach used have been
extensively described elsewhere \cite{BRUCE1,WILDING1} and so will not be
repeated here.  We merely quote the results: $\tilde{T}_c^{lg}=1.024(2),
\tilde{\rho}_c^{lg}=0.327(2), \tilde{\mu}_c^{lg} =-2.767(4)$.

\begin{figure}[h]
\setlength{\epsfxsize}{14.0cm}
\centerline{\mbox{\epsffile{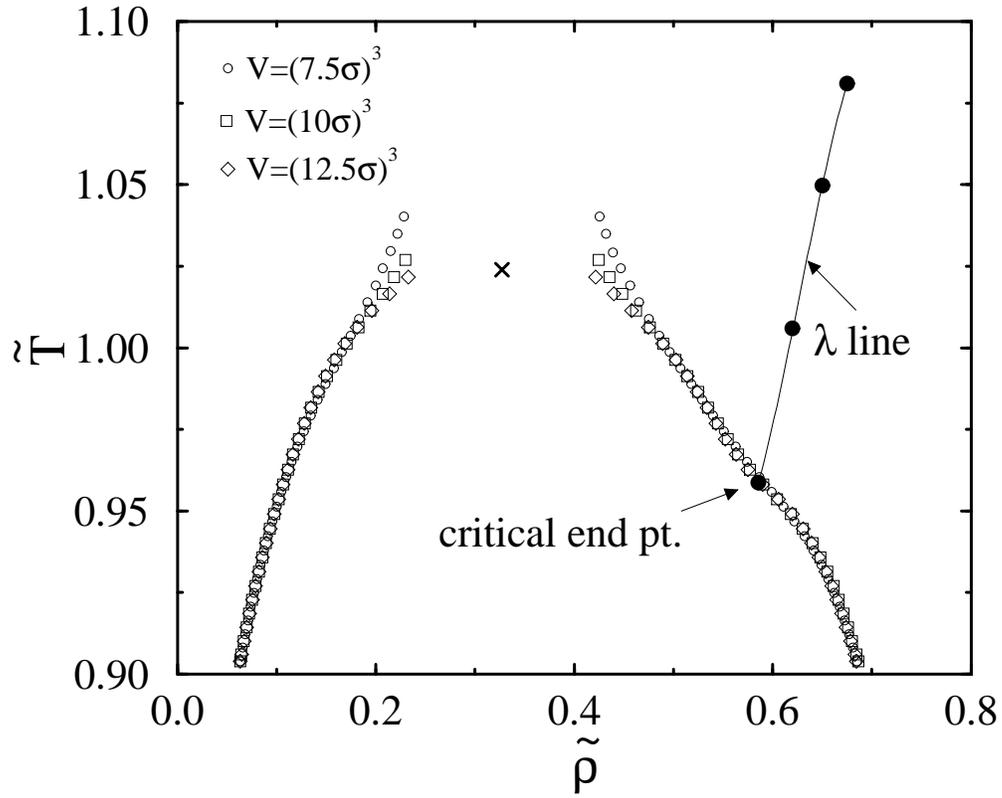}}} 

\caption{The average peak densities corresponding to the coexistence form of
$p_L(\tilde\rho)$ for the three systems sizes studied, plotted as a function of
temperature.  Also shown is the estimated locus of the $\lambda$-line
(filled circles) and the liquid-gas critical point (cross).  Statistical errors do
not exceed the symbol sizes.}

\label{fig:cxcurve}
\end{figure}

In figure~\ref{fig:cxcurve} we present the estimated phase diagram of our
system. Plotted are the coexistence densities for the liquid and gas branches,
the position of the liquid-gas critical point, the locus of the $\lambda$-line
and the position of the critical end point.  One observes that close to the
liquid-gas critical point, strong finite-size effects occur in the peak
densities of $p_L(\tilde\rho)$. These have also been observed and discussed in
the context of a previous study of the pure LJ fluid \cite{WILDING1}.  More
striking, however, is the appearance of a pronounced `kink' in the liquid
branch density close to the critical end point. This is a consequence of the
singular behaviour on the critical line. As one would expect, the gas branch
displays no such kink due to the analyticity of $G_g(\mu,T)$ at $T_e$.

\begin{figure}[h]
\setlength{\epsfxsize}{11.0cm}
\centerline{\mbox{\epsffile{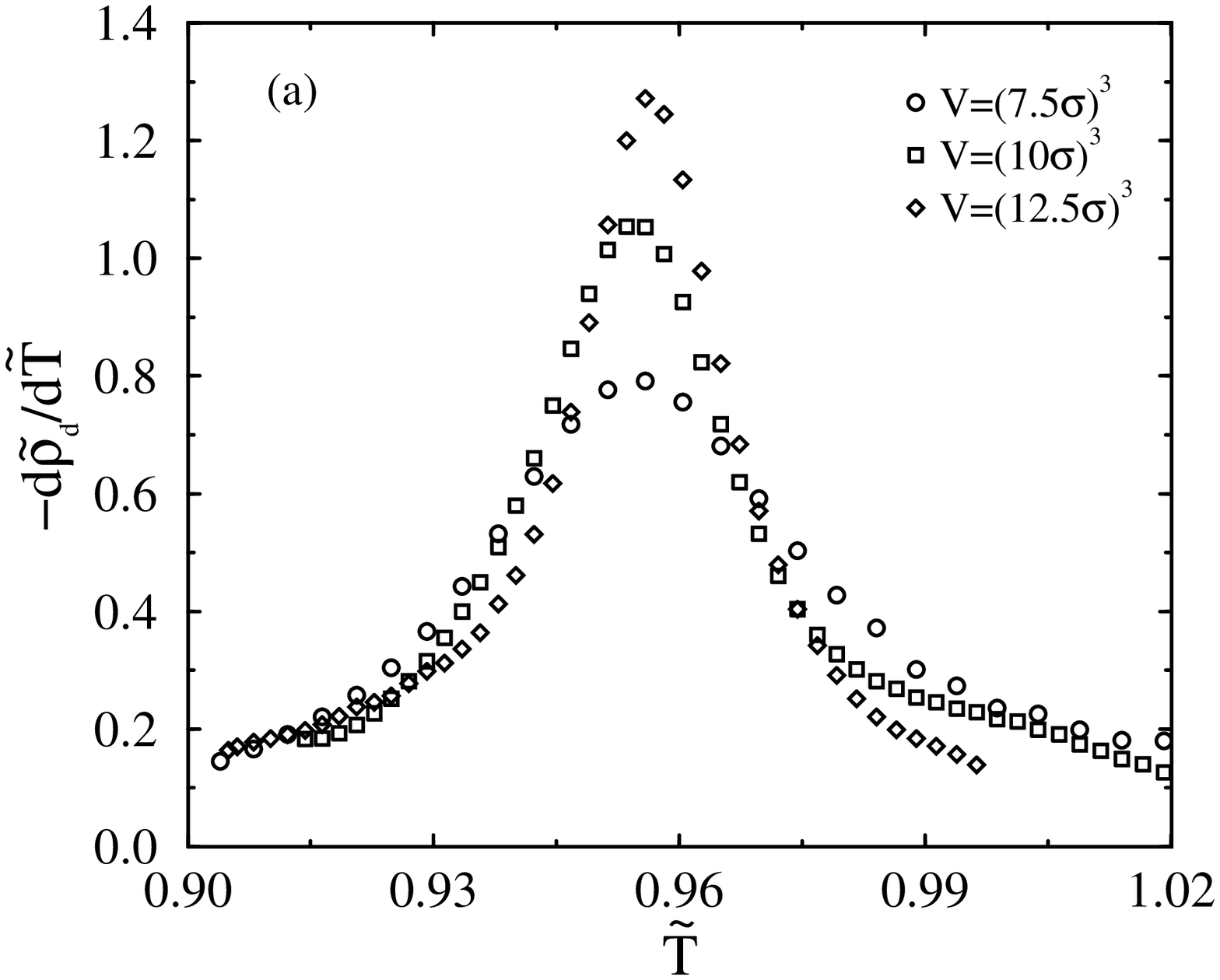}}} 
\centerline{\mbox{\epsffile{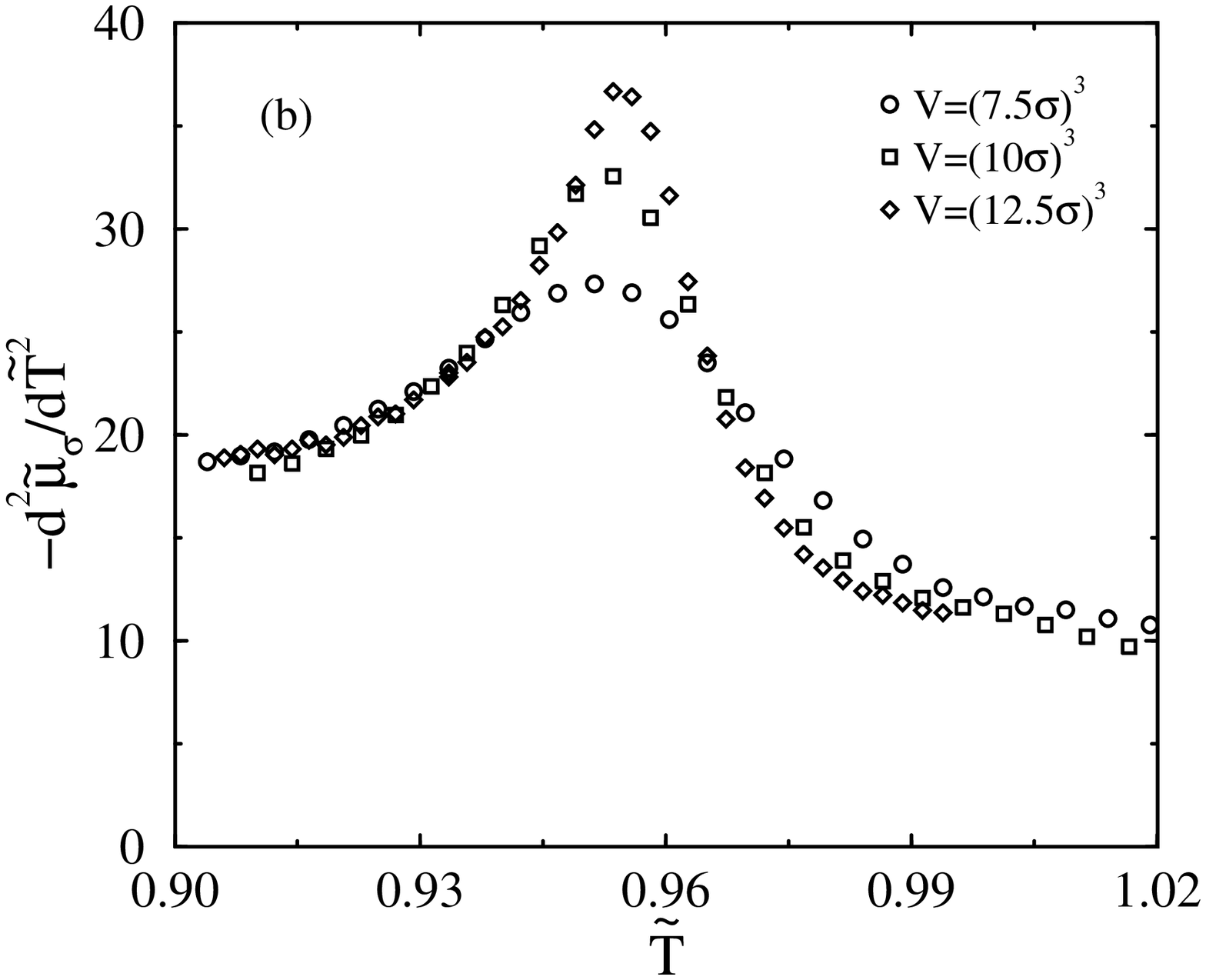}}} 

\caption{{\bf (a)} The numerical temperature derivative of the measured
coexistence diameter $-d\rho_d/d\tilde T$ in the vicinity of the critical end
point temperature. {\bf (b)} The measured curvature of the phase boundary
$-d^2\tilde\mu_\sigma/d\tilde T^2$ in the vicinity of the critical end point
temperature.  In both cases data are shown for the three system sizes
studied.}

\label{fig:diam}
\end{figure}

To probe more closely the behaviour of the coexistence density, we plot in
figure~\ref{fig:diam}(a) the diameter derivative $-d\tilde\rho_d(T)/d\tilde T$, for
the three system sizes studied.  The data exhibit a clear peak close to $T_e$,
that narrows and grows with increasing system size.  Very similar behaviour is
also observed in the curvature of the spectator phase boundary
$-d^2\tilde\mu_\sigma/ d\tilde T^2$, see figure ~\ref{fig:diam}(b).  These
peaks constitute, we believe, the finite-size-rounded forms of the divergences
eqs.~\ref{eq:ddiv} and \ref{eq:cdiv}.  On the basis of finite-size scaling
theory \cite{BINDER}, the peaks are expected to grow in height like
$L^{\alpha/\nu}$, with $\nu$ the correlation length exponent.  Unfortunately,
it is not generally feasible to extract estimates of $\alpha/\nu$ in this way
(even for simulations of lattice Ising models), because to do so requires the
ability to measure the analytic background, for which the present system sizes
are much too small.  Nevertheless, the correspondence of the peak position
with the independently estimated value of $\tilde T_e$, as well as the
narrowing and growth of the peak with increasing $L$ constitutes strong
evidence supporting the existence of the predicted singularities.

\section{Discussion}

\label{sec:disc}

In summary, we have employed grand canonical Monte-Carlo simulations in
conjunction with multicanonical preweighting and histogram reweighting to
study the first order phase boundary near the critical end point of a
continuum binary fluid model.  The results provide clear evidence for the
existence of singularities in the phase boundary chemical potential and the
coexistence curve diameter. They thus constitute strong corroboration of
the scaling arguments on whose basis the singularities are predicted.

Although no experimental observations of singularities at critical end points
have yet been reported, we believe that given an appropriately chosen system
their presence should be relatively easy to detect. In this respect, a
binary fluid model might be a good candidate system for study, since generally
speaking rather precise liquid-gas coexistence curve data can be obtained. In
this case the r\^{o}le of the chemical potential in equation~\ref{eq:csing}
is replaced by the pressure, but otherwise the form of the singularity
remains unaltered. Of course in a real system, the lack of finite-size
rounding should render the singularities more conspicuous than in a
simulation. Moreover, since real binary fluids do not generally possess a
special symmetry between the two fluid components, the chemical potential and
temperature feature in the scaling field $\hat{h}$, changing the
$|t|^{1-\alpha}$ diameter singularity into a much stronger $|t|^\beta$ singularity
(cf.  section~\ref{sec:theory}). In future work we also intend to study the
consequences of this symmetry breaking, via simulations of an asymmetrical
binary fluid model.

\subsection*{Acknowledgements}

The author thanks K.  Binder and D.P.  Landau for stimulating discussions and
encouragement. Helpful correspondence with A.D. Bruce, M.E. Fisher, M. Krech
and M. M\"{u}ller is also gratefully acknowledged. This work was supported
by BMBF project number 03N8008 C.

\newpage

\newpage

\newpage

\newpage

\newpage

\newpage

\newpage

\newpage

\newpage

\end{document}